\def\yes{yes }
 \read-1 to\yesno
\ifx\yesno\yes 
\font\tendl=msym10  scaled \magstep1
\font\sevendl=msym7 scaled \magstep1
\font\fivedl=msym5 scaled \magstep1
\newfam\dlfam \def\dl{\fam\dlfam\tendl}
\textfont\dlfam=\tendl \scriptfont\dlfam=\sevendl
\scriptscriptfont\dlfam=\fivedl
\else
\let\dl=\bf
\fi
\def\versionno{ 4 3.12.93}\def\prepnr{NIKHEF-H/93-28}

\def\ifundefined#1{\expandafter\ifx\csname#1\endcsname\relax}
\makeatletter \ifundefined{new@mathgroup} {} \else
    \new@mathgroup\sffam
    \new@internalmathalphabet\mathsf\sffam{cmss}{m}{n}
    \def\psf{\fontfamily\sfdefault \fontseries\default@series
        \fontshape\default@shape\selectfont\mathsf}
\fi

\def\citen#1{\if@filesw \immediate\write \@auxout {\string\citation{#1}}\fi%
\@tempcntb\m@ne \let\@h@ld\relax \def\@citea{}%
\@for \@citeb:=#1\do {\@ifundefined {b@\@citeb}%
    {\@h@ld\@citea\@tempcntb\m@ne{\bf ?}%
    \@warning {Citation `\@citeb ' on page \thepage \space undefined}}%
    {\@tempcnta\@tempcntb \advance\@tempcnta\@ne
    \setbox\z@\hbox\bgroup\ifcat0\csname b@\@citeb \endcsname \relax
    \egroup \@tempcntb\number\csname b@\@citeb \endcsname \relax
    \else \egroup \@tempcntb\m@ne \fi \ifnum\@tempcnta=\@tempcntb
    \ifx\@h@ld\relax \edef \@h@ld{\@citea\csname b@\@citeb\endcsname}%
    \else \edef\@h@ld{\hbox{--}\penalty\@highpenalty
    \csname b@\@citeb\endcsname}\fi
    \else \@h@ld\@citea\csname b@\@citeb \endcsname \let\@h@ld\relax \fi}%
\def\@citea{,\penalty\@highpenalty\hskip.13em plus.13em minus.13em}}\@h@ld}
\def\@citex[#1]#2{\@cite{\citen{#2}}{#1}}%
\def\@cite#1#2{\leavevmode\unskip\ifnum\lastpenalty=\z@\penalty\@highpenalty\fi%
  \ [{\multiply\@highpenalty 3 #1%
  \if@tempswa,\penalty\@highpenalty\ #2\fi}]}   %
\makeatother 

\newcommand{\ad}[1] {\mbox{ad$_{#1}$}}
\def\aft    {algebraic field theory}
\def\alg    {algebra}

\def\apo    {\mbox{$S$}}
\newcommand{\B}[1]  {\mbox{${\cal B}_{#1}$}}
\def\barhrep{\mbox{$\overline{\varpI}$}}
\def\be     {\begin{equation}}
\def\bfe    {{\bf1}}
\def\bl     {\mbox{$b_{\rm l}$}}

\def\br     {\mbox{$b_{\rm r}$}}
\def\ca     {\mbox{$\cal A$}}
\def\calbh  {\mbox{${\cal B}({\cal H})$}}
\def\calf   {\mbox{$\cal F$}}
\def\calh   {\mbox{$\cal H$}}
\def\cat    {category}
\def\cft    {conformal field theory}
\def\cfts   {conformal field theories}
\def\cgc    {Clebsch\hy Gordan\ coefficient}
\def\cirC   {\!\circ\!}
\def\cj     {^{\scriptscriptstyle +}}
\let\cli=\centerline
\def\coa    {\Q}
\def\coc    {\R}
\def\cop    {\mbox{$\!\!\begin{array}{c}\circ \\[-1.26 em]
            \mbox{\small $\bigtriangleup$}  \end{array}\!\!$}}
\def\coP    {\mbox{$\!\!\!\begin{array}{c}\circ \\[-1.26 em]
            \mbox{\small $\bigtriangleup$}  \end{array}\!\!\!$}}
\def\copp   {\mbox{$\!\!\begin{array}{c}\circ \\[-1.26 em]
            \mbox{\small $\bigtriangleup$}  \end{array}\!\!\!'$}}
\def\copP   {\mbox{$\!\!\!\begin{array}{c}\circ \\[-1.26 em]
            \mbox{\small $\bigtriangleup$}  \end{array}\!\!\!'\!$}}
\def\Co     {{\dl C}}
\def\complex{{\dl C}}
\newcommand{\complx}[2] {\mbox{{\dl C}\raisebox{#1 em}{$\scriptstyle #2$}}}
\def\cou    {\mbox{$\epsilon$}}
\def\cst    {{\rm C}$^*$}
\def\cstalg {C$^*$-algebra}
\def\DHR    {Dop\-licher\hy Haag\hy Roberts\ }
\def\doro   {Dop\-licher\hy Roberts\ }
\newcommand{\e}[3] {\mbox{$e^{#2,#3}_{#1}$}}
\def\ee     {\end{equation}}

\def\emt    {ener\-gy-mo\-men\-tum-ten\-sor}
\newcommand{\erf}[1] {(\ref{#1})}
\newcommand{\erF}[1] {(\ref{#1})}

\def\findim {fi\-nite-dimen\-sio\-nal}

\def\furu   {fusion rule}
\def\futnote#1 {\footnote{~#1}\ }

\def\h      {\mbox{$H$}}
\let\H=\h

\def\hilsp  {Hilbert space}
\def\hilsph {Hilbert space \calh}

\def\hrep   {\mbox{$\varpI$}}

\def\hy     {$\mbox{-\hspace{-.66 mm}-}$}
\def\id     {{\sl id}}
\def\ii     {{\rm i}}
\def\infdim {infi\-nite-dimen\-sio\-nal}
\def\irrep  {irreducible representation}
\def\ite    {\item[$\triangleright$]}
\def\itE    {\item[{\tiny$\Diamond$}]}
\def\kma    {Kac\hy Moody algebra}
\newcommand{\labl}[1] {\label{#1}\ee}
\def\llb    {\mbox{\Large(}}

\def\loplus {\mbox{\Large $\oplus$}}
\def\lrb    {\mbox{\Large)}}
\def\mU     {\mbox{$\mu$}}
\def\muhrep {\mbox{$\mu^{}_{\varpI}$}}
\def\mul    {\mbox{\sc m}}
\def\npqr   {\mbox{$N_{pq}^{\ \ r}$}}
\let\oeta=\eta
\def\one    {\mbox{{\small 1}$\!\!$1}}
\def\onedim {one-dimen\-sio\-nal}
\def\oti    {\times}

\def\otim   {\otimes}

\def\pefro  {Perron\hy Fro\-benius\ }
\def\Q      {\mbox{$Q$}}
\def\QQ     {\mbox{$\hat Q$}}
\def\qft    {quantum field theory}
\def\qfts   {quantum field theories}

\def\qg     {quantum group}
\def\qs     {quantum symmetry}
\def\qss    {quantum symmetries }
\long\def\query#1{\hskip 0pt{\vadjust{\everypar={}\small\vtop to 0pt{\hbox{}%
     \vskip -13pt\rlap{\hbox to 46.3pc{\hfil{\vtop{\hsize=8pc\tolerance=6000%
     \hfuzz=.5pc\rightskip=0pt plus 3em\noindent#1}}}}\vss}}}}%

\newcommand{\RR}[3] {\mbox{$R^{(#1 #2)_{#3}}$}}
\def\reals  {{\dl R}}
\def\R      {\mbox{$R$}}

\def\rep    {represen\-ta\-ti\-on}
\def\repcat {representation category}
\def\rha    {rational Hopf algebra}
\def\rphys  {\mbox{{\small$\Pi$}$_{\rm phys}$}}
\def\rR     {{\rm R}}
\def\RR     {\mbox{$\hat R$}}
\newcommand{\Sy}[1]  {\mbox{${\cal S}_{#1}$}}
\def\ssi    {semisimple}
\def\stalg  {$^*$-algebra}
\def\statdim{statistical dimension}
\def\susE   {superselection\ }
\def\suse   {superselection sector}
\def\sym    {symmetry}
\def\syms   {symmetries}
\def\takrein{Tan\-naka\hy Krein\ }
\def\tims   {\times}
\def\tqg    {truncated quantum group}
\def\twodim {two-dimen\-sio\-nal}
\def\UN     {\mbox{$e$}}
\def\ut     {\mbox{${\cal U}_{\rm t}$}}
\def\vac    {\mbox{$\Omega$}}
\def\varpI  {\mbox{$\varpi$}}
\def\veps   {\mbox{$\varepsilon$}}
\def\vept   {\mbox{$\varepsilon$}}
\newcommand{\version}[1] {}

\def\vna    {von Neumann algebra}
\def\wrt    {with respect to\ }

\def\zet    {{\dl Z}}

\documentstyle[11pt]
{article}

\ifx\yesno\yes 
\else  \fi

\begin{document} \version{\versionno}
\vskip 5mm
   \rightline{{\sf\prepnr}}
     \vskip 1 mm \rightline{{\sf hep-th/9312026}} \vskip 14mm
     \cli{\bf\Large THE\,\, QUANTUM\,\, SYMMETRY}\vskip 2.3 mm
     \cli{{\bf\Large OF RATIONAL FIELD THEORIES} \raisebox{3 mm}{${\$}$} }
    \vskip 14 mm \cli{\sc {}~J\"urgen Fuchs $^{\&}$}\vskip 2 mm{}%
     \cli{NIKHEF-H}\vskip .5 mm \cli{Kruislaan 409}\vskip .5 mm
     \cli{NL -- 1098~SJ~~Amsterdam} \vskip 11mm
     \begin{quote}{\bf Abstract.}\ \,
The quantum symmetry of a rational \qft\ is a \findim\ multi-matrix algebra.
Its \repcat, which  determines the \furu s and braid group \rep s of
\suse s, is a braided monoidal \cst-category. Various properties of
such algebraic structures are described, and some ideas concerning the
classification programme are outlined.
     \end{quote}
   \vfill --------------------- \\ $^{\$}$~~%
Invited talk given at the III. International Conference on Mathematical
Physics, String\\ \mbox{~~} Theory and Quantum Gravity, Alushta, Ukraine,
June 1993; to appear in Teor.Mat.Fiz. \\[1 mm]
$^{\&}$~~Heisenberg fellow

\section{Symmetries and observables}

The notion of symmetry is one of the most fundamental concepts in
science.
An {\em internal\/} \sym\ leaves observable quantities invariant.
It only changes the particular manner
in which a physical system is described and correspondingly
represents a certain amount of redundancy. Therefore it is often
a good idea to eliminate internal \syms\
so as to arrive at a description in terms of observables only.
But there are also various situations where, on the contrary, the redundancy
is highly welcome. For instance, often the use of redundant variables
leads to simpler coordinates and thereby facilitates a perturbative
treatment. Another example arises in the context of
what I will call {\em \qs}. To give a first impression of what is
meant by this term, let me specialize to the specific case of \cft:
By a \qs\ of a \twodim\ \cft\ I mean an internal \sym\ whose \rep\
theory reproduces the basis independent contents of the operator product \alg,
i.e.\ the \furu s, and which is compatible with the duality properties
of chiral blocks. A model independent characterization of the notion
of \qs\ will be given at the beginning of section \ref{qs}.

While conformal invariance is operational in the latter characterization,
it is far from being an essential ingredient. Rather, the type of \qss
encountered in \cft\ turns out
to be relevant to relativistic \qft\ in general.
To make this remark more concrete, let me
consider a specific axiomatic formulation of \qft, namely
the algebraic theory of \suse s \cite{dohr,dohr4,Kamr,HAag}, to which for
short I will refer in the sequel by the term {\em \aft}.
In the \aft\ framework a theory is described
in terms of nets of \vna s $\ca({\cal O})$
which are indexed by specific open
subsets $\cal O$ of Minkowski space-time. Consequently, the proof of
various results from \aft\ on which I will rely below requires the
mathematics of \vna s and subfactors \cite{SUnd,muvo2,jone,GOdj};
however, the essence of these results is in all relevant cases rather
plausible already from a more heuristic point of view.  My perspective
will therefore be to take these results for granted and investigate their
implications for \qs.

Classical physics can be described via a configuration space $X$
endowed with a measure $\mu$; the dynamics is then described in terms
of the elements of the \alg\ ${\cal L}^\infty(X,\mu)$ of bounded
measurable functions on $X$.
In quantum physics, one trades the configuration space,
respectively the commutative algebras of functions on it,
for non-commutative \stalg s of operators.
The analogue of the measures $\mu$ are states (normalized positive
linear forms) over the \alg. A state induces a scalar product
on the \alg, and the completion of the algebra \wrt the associated
norm is a separable \hilsph\ on which the algebra acts by multiplication.
Thus physical states correspond to the vectors $\psi$
of a \hilsph; the analogue of ${\cal L}^\infty(X,\mu)$
is then the field algebra \calf\ which is a sub\alg\ of the \alg\
\calbh\ of bounded operators on \calh.
The vectors of \calh\ can be thought of as being
created by acting with field operators $f\in\calf$ on a vacuum vector
\vac. In this context,
the observables are the (self-adjoint) elements of a $^*$-subalgebra
\ca\ of \calf. Any measuring apparatus is contained in a
bounded region $\cal O$ of space-time, and hence there must exist local
observable \alg s $\ca({\cal O})$ of observables measurable in
$\cal O$. The local observable algebras associated to
causally disconnected regions commute among
one another (Einstein causality).
The total observable \alg\ $\ca\subseteq\calf$ is the quasilocal
\cstalg\ $\overline{\bigcup_{\cal O}\ca({\cal O})}$\,.

A distinctive feature of \qss in low-dimensional field theory is that
they correspond to {\em \findim\/} algebras. Recall that in classical
physics the (finite) symmetry transformations correspond to the elements
of a group $G$, or rather to the associated group algebra $\complex G$.
If $G$ is infinite, say a Lie group, then $\complex G$ is \infdim.
As it turns out, this situation prevails in quantum theory as
long as the dimensionality of space-time is large enough. In contrast,
in low-dimensional \qft\ there are special systems, the so-called
rational field theories, for which the \qs\ is a \findim\
\alg\ which is generically not a group \alg.
Heuristically, the possibility of having more general structures
than group \alg s can be understood by investigating the question of
what happens to the group of symmetries in the course of
quantization of a classical system. Namely, elements of the group $G$
can be viewed as acting on the points of the configuration space.
Since in quantum physics the configuration space is no longer present,
the group elements are not needed any more either.

\section{Superselection sectors}

In algebraic field theory the observables are taken as the
basic objects. Because of Einstein causality, observables must
commute at space-like separations, or in other words, the statistics of
observables is bosonic. One of the challenges of algebraic \qft\
is the investigation of the possible statistics (i.e., the behavior
\wrt permutations) of non-observable fields.
The action of the field algebra \calf\ on the vacuum vector \vac\ is
generically reducible; correspondingly, there are
sub-\hilsp s of the \hilsph\ that are orthogonal to each other.
These subspaces are called {\em \suse s}; denoting them by
$\calh_\alpha$, one has the sector decomposition
  \be  \calh=\loplus_\alpha\calh_\alpha \,.   \labl{sec}
Observables act within the sectors.
(This implies that the relative phases of vectors belonging to different
subspaces ${\cal H}_\alpha$ cannot be observed. Hence, in contrast to
the situation in quantum mechanics, the superposition
principle is not valid universally, but holds only within sectors.)
Thus
each sector $\calh_\alpha$ carries a \rep\ $\pi_\alpha$ of \ca.
The general \rep\ theory of \ca\ is rather wild, but fortunately
in \qft\ only a restricted class of \rep s is relevant. The precise
requirements that these physical \rep s must meet depend on the
specific axiomatic framework; in the present context, the relevant notion
\cite{dohr,dohr4,Kamr,HAag} is the one of {\em DHR} ({\em \DHR}$\!$) -\rep s.
Recall that a vacuum \rep\ $\pi_0$ of \ca\
is a positive-energy \rep\ for which the associated
\hilsp\ $\calh_0$ contains a unique (up to normalization) vector \vac,
called the vacuum vector, which is cyclic and separating for \calh\
and is invariant under the relevant
space-time symmetry transformations, in particular under translations.
Then by definition a DHR \rep\ is a \rep\ which is
isomorphic to $\pi_0$ outside some bounded region, or
in other words, is a `local excitation of a vacuum \rep'.
It will also be assumed below that the index of inclusion of the \alg\
$\pi_\alpha(\ca({\cal O}))$ in the commutant of $\pi_\alpha(\ca({\cal O}'))$
(${\cal O}'$ denotes the space-like complement of ${\cal O}$)
\wrt \calbh\ is finite. DHR \rep s with this property are said to have
{\em finite statistics}.

In more technical terms, DHR \rep s $\pi_\alpha$ are characterized by
being unitarily equivalent to $\pi_0$ in the sense that
  \be  \pi_\alpha \cong \pi_0\circ\rho_\alpha \,, \labl{pp}
where $\rho_\alpha\!:\,\ca({\cal O})\rightarrow\ca({\cal O})$
are endomorphisms of the local algebras $\ca({\cal O})$ which
act as the identity map on the \vna\ generated by the
local observables in ${\cal O}'$.
(This implies that the \alg s generated by the local observables in
different sectors are isomorphic, and the sectors can only be
distinguished by global quantities, which are referred to as `\susE charges'.)

Several properties of DHR \rep s with finite statistics are relevant to
the investigation of \qs. They are given in the following list.

\begin{itemize}

\ite The \rep s are irreducible, and they appear in the sector decomposition
\erf{sec} with finite multiplicity. Accordingly \erf{sec} can be rewritten as
  \be  \calh=\loplus_p(\calh_p\oti\complx{.54}{n_p}) \,,   \labl{seC}
where the \rep s $\pi_p$ corresponding to the \hilsp s $\calh_p$ are
pairwise inequivalent, and where $n_p<\infty$ are non-negative integers.
\\ The vacuum sector is non-degenerate, $n_0=1$.

\ite It is possible to define a tensor product of
(unitary equivalence classes of) the \rep s $\pi_p$ of \ca; namely,
  \be  \pi_p\times\pi_q \cong \pi_0\circ\rho_p\circ\rho_q  \,. \labl{pipi}

\ite The product of sectors is completely reducible. Thus
  \be \pi_p\times\pi_q \cong \loplus_r(\complx{.59}{N_{pq}^{\ \;r}}
  \oti\pi_r)  \labl{rhorho}
for some non-negative integers \npqr. I will refer to these
integers  as \furu\ coefficients. One has $\pi_0\oti\pi_q\cong\pi_q$,
i.e. $N_{0q}^{\ \;r}=\delta_q^{\;r}$.

\ite To any \rep\ $\pi_p$ there is associated a unique
\rep\ $\pi_{p\cj}$, called {\em conjugate \/} to $\pi_p$,
such that the product of $\pi_p$ with $\pi_{p\cj}$
contains $\pi_0$, and then $\pi_0$ appears precisely once in this
product. In terms of \furu\ coefficients, $N_{pq}^{\; \ 0}=\delta_{p,q\cj}^{}$.

\ite The composition of the endomorphisms $\rho_p$ of \ca\ is
associative. Further, while the composition of endomorphisms is in
general not commutative, it is still commutative up to unitary equivalence.
As a consequence, the tensor product of DHR \rep s is associative up to
isomorphism and commutative up to isomorphism. For the fusion coefficients,
this implies $ N_{pq}^{\ \;s} N_{sr}^{\ \;t}= N_{qr}^{\ \;s} N_{ps}^{\ \;t}$
and $N_{pq}^{\ \;s}=N_{qp}^{\ \;s}$.

\ite The matrices $N_p$ with entries $(N_p)_q^{\ r}=\npqr$ are
simultaneously diagonalized by a symmetric matrix $S$ \cite{rehr4}.
In general, this matrix may be degenerate, but in that case it is
possible to enlarge \ca\ such that the diagonalization matrix arising
for the enlarged observable algebra is non-degenerate, and hence can
be chosen to be unitary.
 \end{itemize}

Further, the statistical properties of \suse s may be summarized as
follows. For any natural number $m$, each sector $\pi_p$
carries an irreducible matrix \rep\ of the braid group \B m on $m$
strands. These \rep s
are to a large extent characterized by two numbers, namely a by
a phase $\oeta_p$ and by a positive real number $d_p$;
in the limiting case where the \rep s are in fact \rep s of the
symmetric group $\Sy m\subset\B m$ (permutation group statistics), these
numbers
correspond to the sign $\pm1$ distinguishing bosons from fermions and
to the dimensionality of the \Sy m-\rep, respectively. With respect to
statistics, the distinctive feature of \aft\ is that
statistics is an {\em intrinsic\/} property of \suse s, in the sense that
one can directly describe the numbers $d_p$ and $\eta_p$ in terms of the
sectors. Recall that in quantum theory
the statistics of fields is encoded in their commutation relations
for space-like separation. In the \aft\ framework, this information is
described by the so-called {\em statistics parameters\/} of the sectors.
The inverse modulus of the statistics parameter, called the {\em
\statdim}, reproduces the number $d_p$, and the phase of the
statistics parameter, called the {\em statistics phase}, coincides with
the phase $\oeta_p$. The physical interpretation of the \statdim\ is that
it measures the deviation from Haag duality, i.e.\ from
the property $\pi_p(\ca({\cal O'})')= \pi_p(\ca({\cal O}))$,
where the prime on the \alg\ denotes the commutant \wrt \calbh; more
precisely,
  \be  d_p^2={\rm Ind}[ \pi_p(\ca({\cal O'})') \,:\,
  \pi_p(\ca({\cal O})) ] \,, \labl{ind}
is the index
of the inclusion of $\pi_p(\ca({\cal O}))$ in $\pi_p(\ca({\cal O}')')$.
Without essential loss of generality, the vacuum sector can be assumed
to be Haag dual, and hence $d_0=1$. I will do so, and correspondingly
identify \ca\ with its vacuum \rep\ $\pi_0(\ca)$.

One can actually compute the statistics parameter of a sector $\pi_p$
rather directly from the associated endomorphism $\rho_p$. Namely,
one has $\Phi_p(\veps_{pp})=(\oeta_p/d_p)\,\bfe$. Here $\veps_{pq}$
is the `statistics operator' which describes
the noncommutativity of the composition of the endomorphisms $\rho_p$
and $\rho_q$, and $\Phi_p$ is a `left-inverse' of $\rho_p$, i.e.\
a positive linear map with the property that $\Phi_p
\circ\rho_p$ is the identity map on $\cal O$.
Further, the index \erf{ind} of \vna s coincides with the index
  \be  {\rm Ind}[ \ca({\cal O}) \,:\, \rho_p(\ca({\cal O}))]  \labl{ind'}
of \cstalg s.

\vskip 4 mm

In the special case that one is actually dealing with a (\twodim)
{\em conformal\/} \qft, it is straightforward
to make contact to the framework of the bootstrap formulation
of these theories. The chiral half of a (unitary) \cft\ corresponds to a
\qft\ on a circle S$^1$ which is a compactified light ray of \twodim\
Minkowski space. At a heuristic level, one then has the following
correspondences (compare \cite{jf20};
for a more detailed analysis, see \cite{frrs2,gafr,brgl,vecs,szla}):
 \begin{itemize}
\ite The elements of the local observable algebras $\ca({\cal O})$
correspond to bounded functions of operators of the form
$\oint_{{\rm S}^1_{}}{\rm d}z\,f(z)W(z)$, where $f$ is a test function with
support in $\cal O$ and $W$ an element of the enveloping algebra of
the chiral symmetry algebra $\cal W$.
\ite The \suse s correspond to the physical \rep s of $\cal W$, or in
other words, to the chiral families (headed by primary fields) of the
theory.
\ite The vacuum sector corresponds to the algebra $\cal W$ itself,
i.e.\ to the identity primary field.
\ite The \statdim\ of a sector coincides with the quantum dimension
of the associated primary field, and the statistical phase with
exp$(2\pi\ii\Delta)$, with $\Delta$ the conformal dimension of the
primary.
\ite The rationality property means that the number of primary
fields is finite.
\ite The composition of sectors corresponds to the \furu s of the \cft\
which describe the basis independent contents of the operator product
\alg\ (recall that the \furu s must not be mixed
up with the ordinary tensor products of \rep s of the chiral \alg\
$\cal W$).
\ite Choosing the observable algebra such that the matrix which diagonalizes
the \furu s is symmetric and unitary is equivalent to choosing the maximally
extended chiral algebra, i.e.\ incorporating all purely holomorphic
primary fields with integer conformal dimension into $\cal W$.
 \end{itemize}

\section{Quantum symmetry} \label{qs}

The above considerations motivate the following terminology.
 \begin{quote} A\, {\em q\underline{uantum symmetr}y\/}
\,is a symmetry structure \h\ which allows for an intrinsic \\[.6mm]
determination
of the \furu s and of the statistics of \suse s. \end{quote}
In view of the results presented in the previous section, one of
the characteristic properties of a \qs\ is thus that the \rep\ theory of
the observable \alg\ \ca\ should coincide with the \rep\ theory
of the \qs\ \h. In more mathematical terms,
the physical \rep s of \ca\ can be understood in terms of
the category of \rep s of \h.
(This is sometimes rephrased by saying that the \repcat\ of \h\
should be isomorphic to the category of physical \rep s of \ca.
One should keep in mind, however, that according to standard terminology
the objects of a \repcat\ are {\em \findim\/}
matrix \rep s; thus in the context of \ca\ which has no
faithful \findim\ \rep s, this use of category theoretic terms is
a bit non-standard.)
As already mentioned, the qualification `physical' for the \rep s of \ca\
is essential. Its precise meaning depends on the concrete physical
situation one is interested in; in the present context,
a physical \rep\ is a DHR \rep\ with finite statistics.

By construction, the `gauge charge' structure describing the \rep\
labels \wrt the \qs\ \h\
is thus in one-to-one correspondence with the \susE structure of the theory,
and it is natural to identify the two concepts of charge \cite{rehr6}.
In terms of the \hilsph, this implies that the
sector decomposition \erf{sec} gets generalized to \erf{seC}, or
more precisely, to
  \be  \calh=\loplus_p(\calh_p\times V_p) \,,  \labl{hv}
where $V_p\cong\complx{.54}{n_p}$ are irreducible \h-modules,
while in terms of operator algebras, one has
  \be  \calbh\supset\ca\times H. \labl{sa}

{}From the properties of the tensor product of \suse s it follows in
particular that the \repcat\ of \h\ must have the properties of
a {\em braided monoidal \cst-category}. These properties include in particular
the commutativity of certain diagrams describing
the composition of isomorphisms among tensor products which are
calculated in different orders. Among these, the most important ones
are the pentagon and hexagon relations, whose description in terms of the
\qs\ \h\ will be given in \erf{pen} -- \erf{hex2} below; in terms of
commutative diagrams, they look as follows:\\[1 mm]
  {\small
 $${\rm Pentagon:}\qquad \begin{array}{ccccc} &&
\fbox{$\pi_1\oti((\pi_2\oti\pi_3)\oti\pi_4)$} & \longrightarrow &
\fbox{$(\pi_1\oti(\pi_2\oti\pi_3))\oti\pi_4$} \\[.3 mm]
& \nearrow  \\
\fbox{$\pi_1\oti(\pi_2\oti(\pi_3\oti\pi_4))$} &&&& \downarrow \\[.3 mm]
& \searrow  \\
&& \fbox{$(\pi_1\oti\pi_2)\oti(\pi_3\oti\pi_4)$} & \longrightarrow &
\fbox{$((\pi_1\oti\pi_2)\oti\pi_3)\oti\pi_4$} \\[3 mm]
\end{array} $$ \vskip 5mm
 $${\rm First\ hexagon:} \hskip 5.2em
\begin{array}{ccccc}
\fbox{$\pi_1\oti(\pi_2\oti\pi_3)$} & \longrightarrow &
\fbox{$\pi_1\oti(\pi_3\oti\pi_2)$} & \longrightarrow &
\fbox{$(\pi_1\oti\pi_3)\oti\pi_2$} \\[3 mm]
\downarrow &&&& \downarrow \\[3 mm]
\fbox{$(\pi_1\oti\pi_2)\oti\pi_3$} & \longrightarrow &
\fbox{$\pi_3\oti(\pi_1\oti\pi_2)$} & \longrightarrow &
\fbox{$(\pi_3\oti\pi_1)\oti\pi_2$} \\[3 mm]
\end{array} \hskip 3em $$ \vskip 3mm
 $${\rm Second\ hexagon:} \hskip 4.1em
\begin{array}{ccccc}
\fbox{$(\pi_1\oti\pi_2)\oti\pi_3$} & \longrightarrow &
\fbox{$(\pi_2\oti\pi_1)\oti\pi_3$} & \longrightarrow &
\fbox{$\pi_2\oti(\pi_1\oti\pi_3)$} \\[3 mm]
\downarrow &&&& \downarrow \\[3 mm]
\fbox{$\pi_1\oti(\pi_2\oti\pi_3)$} & \longrightarrow &
\fbox{$(\pi_2\oti\pi_3)\oti\pi_1$} & \longrightarrow &
\fbox{$\pi_2\oti(\pi_3\oti\pi_1)$} \\[3 mm]
\end{array} \hskip 3em $$
 } 

In the sequel the main interest will be in {\em rational\/} field theories,
i.e.\ theories for which
the number of sectors is finite. It is then important to stress
that the dimensionalities $n_p$ of \h-modules and the statistical
dimensions $d_p$ are conceptually rather different. To analyze this
distinction, it is convenient to
define a {\em dimension function\/} for the product \erf{rhorho}
to be a map $D:\ p\mapsto D_p\in\reals_+$ satisfying
$D_0=1$ and $D_{p\cj}^{}=D_p$ as well as
  \be  D_p D_q=\sum_r\npqr\,D_r \,.  \labl{dpqr}
For any rational field theory, there is in fact a unique dimension
function $D$; namely, the numbers $D_p$ are just the components
of the common normalized \pefro eigenvector of the fusion matrices
$N_q$. As it turns out \cite{rehr4,FRke}, these components
precisely coincide with the \statdim s $d_p$. Moreover, for many
rational theories one easily deduces from the explicit form of
the \furu s that the components of the \pefro vector are
non-integral \cite{FRke,fuva3,jf18,nill3,krat2,hAri}, from which it
follows in particular that the integral dimensionalities $n_p$
cannot coincide with the dimension function. Rather, while
$n_0=1$ and $n_{p\cj}^{}=n_p$ are fulfilled, one has in general only
the inequality
  \be  n_p n_q\geq\sum_r\npqr\,n_r \,.  \labl{npqr}
Notice that the algebras whose embedding index is given by
\erf{ind'} are \infdim, so that
it is no wonder that the \statdim\ is generically non-integral.

\section{The structure of \h}

By the properties of the \rep s $\pi_p$ of \ca\ listed in section 2,
the following properties of \h\ \cite{vecs} are implied by
the requirement that \h\ be a \qs.
\begin{itemize}

\ite Because of the inclusion relation \erf{sa}, \h\ is a $^*$-subalgebra of
\calbh, and hence is an associative \stalg\ with unit $e$. \\
(This also follows from a more general reasoning. Namely,
assume that, in each \rep, \h\ is generated by
unitary elements $u\in\calbh$ (corresponding to inner
automorphisms ad$_u$). Then the mere fact that for $h,\,h'\in\h$ the product
$u(h)u(h')$ is defined implies via the basic
\rep\ property that there should exist a product $hh'$ of $h$ and $h'$
such that $u(hh')=u(h)u(h')$,
which is associative owing to the associativity of the product in \calf.
Similarly, the existence of a unit and of a $^*$-involution correspond to
the relations $\calf\ni\bfe=u(e)$ and $(u(h))^*=u(h^*)$.)

\ite According to the decomposition \erf{hv}, the \irrep s $\hrep_p$
of \h\ are all \findim. As a consequence, for any rational field
theory, \h\ is a \findim\ \alg.

\ite On the set of \h-\rep s, a tensor product can be defined.
(This is not a trivial fact. Recall that for generic \alg s the
notion of a `product' of \rep s, in contrast to the notion of Kronecker
product of the associated modules, is not well-defined.)
For the \repcat\ this means that it is a monoidal \cat, and for \h\ itself
that it is endowed with a co-multiplication
  \be  \cop: \quad \h\rightarrow\h\times\h. \labl{Cop}
\h\ is a $^*$-bi-algebra, in particular
the coproduct is a $^*$-homomorphism (but not necessarily unital).

\ite The product of \rep s  is completely reducible, with
\furu\ coefficients as in the decomposition \erf{rhorho}.
This means that the \repcat\ and hence also \h\ itself must be \ssi.
In particular, for a rational theory,
\h\ is a \findim\ \ssi\ associative \alg, and hence a multi-matrix \alg,
i.e.\ a direct sum of full matrix \alg s,
  \be  \h=\loplus_p M^{}_{n_p}(\complex) \,.  \labl{hmnp}

\ite \h\ is endowed with a co-unit $\cou : \ H \to\complex,$
which is a unital $^*$-homomorphism.\\
This is because as a symmetry, \h\ leaves the vacuum vector \vac\ invariant,
or more precisely (since not the vectors $\psi\in\calh$, but rather
the rays $\lambda\psi$, $\lambda\in\complx{.5}*$, are the relevant quantities),
it acts on \vac\ by scalar multiplication, $h\cdot\vac=\vac\,\cou(h)$.
Obviously, $\cou (e)=1$, and $\cou(a^*)=\overline{\cou(a)}$ for all $a\in\h$.
It is also immediate that the specific \onedim\ \rep\ of \h\
carried by the vacuum sector is precisely the co-unit.

\ite As a \rep, \cou\ plays the role of the identity element in the
\rep\ ring; thus the product of any \h-\rep\ \hrep\ with the
co-unit is isomorphic to \hrep. This implies the existence of
unitary elements $a_{\rm r}$ and $a_{\rm l}$ of \h\ such that
$(\cou\times\id)\circ\coP=\ad{a_{\rm r}}$,
$(\id\times\cou)\circ\coP=\ad{a_{\rm l}}$. It can be seen that a change
in $a_{\rm r}$ and $a_{\rm l}$ changes \h\ essentially only up to
isomorphism, so that it is no loss of generality to assume
$a_{\rm r}=e=a_{\rm l}$; doing so, the latter relation simplifies to
  \be  (\cou\times\id)\circ\cop=\id=(\id\times\cou)\circ\cop \,.  \labl{eid}

\ite To any \h-\rep\ there exists a unique conjugate \rep.
For \h, this implies the existence of a linear $^*$-anti-automorphism
$\apo\!:\,\h\to\h$, called the antipode, and of
invertible elements $\bl,\, \br\in\h$ satisfying
 $\mul\circ(\id\times\mul')${\large(}$(\id\times\apo)\cirC
  \cop(a)\otim\bl)${\large)}$= \bl\,\cou(a)$ and
 $\mul\circ(\id\times\mul')${\large(}$(\apo\times\id)\cirC
  \cop(a)\otim\br)${\large)}$= \br\,\cou(a)$.
Here \mul\ and \mul$'\equiv\mul\circ\tau$ denote the product and `opposite'
product of \h, respectively ($\tau\equiv\tau_{12}$ stands for the map that
interchanges the tensor product factors).

\ite The tensor product of \h-\rep s is commutative up to isomorphism.
Therefore there is a {\em cocommutator\/} \coc\ intertwining between
\cop\ and the opposite coproduct $\copp\equiv\tau\circ\cop$. The cocommutator
must be an element of $\h\oti\h$. The intertwining property reads
  \be  \copp(a)\cdot\coc=\coc\cdot\cop(a) \labl{coc1}
for all $a\in\h$. A coproduct \cop\ satisfying \erf{coc1}
is said to be quasi-cocommutative.
Further, \coc\ must be almost unitary in the sense that
$\coc\cdot\coc^*=\copp(e)$, $\coc^*\cdot\coc=\cop(e)$, and must satisfy
$\coc\cdot\coc^*\cdot\coc=\coc$.

\ite The tensor product of \h-\rep s is associative up to isomorphism.
Accordingly, there is a {\em coassociator\/} $\coa\in\h\oti\h\oti\h$
intertwining between $(\id\oti\coP)\circ\coP$ and $(\coP\oti\id)\circ\coP$,
  \be  \llb(\coP\times\id)\circ\cop(a)\lrb\cdot\Q= \Q\cdot\llb(\id\times\coP)
  \circ\cop(a)\lrb  \labl{coa1}
for all $a\in\h$. A coproduct \cop\ satisfying \erf{coa1} is said to
be quasi-coassociative. Further, \Q\ must be almost unitary in the
sense that $\Q\cdot\Q^*=(\coP\times\id)\circ\coP(e)$,
$\Q^*\cdot\Q=(\id\times\coP)\circ\coP(e)$, and must satisfy
$\Q\cdot\Q^*\cdot\Q=\Q$.

\ite The coassociator \coa\ and the cocommutator \coc\ satisfy
the compatibility conditions needed for the \repcat\ of \h\ to
constitute a braided monoidal \cst-category. More concretely,
various specific combinations of unitality-,
commutativity-, and associativity-isomorphisms of multiple tensor products
lead to identical (not just up to an isomorphism) results.
\\[1.5 mm] This is expressed by the following set of equations:
\itE the triangle identity, which with the choice \,$a_{\rm r}=e=a_{\rm l}$\,
reads $(\id\oti\cou\oti\id)(\coa) =\cop(\UN)$\,;
\itE the square identities \,$\mul_5\circ\tau_{12}\tau_{45}(\br\otim
(\apo\oti\id\oti\apo)(\coa)\otim\bl)=\UN$\, and\\
$\mul_5\circ\tau_{12}\tau_{45}(\bl\otim(\id\oti\apo\oti\id)(\coa^*)
\otim\br)=\UN$\,;\\[1 mm]
(Here $\mul_\ell$ denotes multiple products, defined inductively by
$\mul_2=\mul$, $\mul_\ell=\mul\circ(\mul_{\ell-1}\oti\id)$.)
\itE the pentagon identity
  \be  (\coP\oti\id\oti\id)(\coa) \cdot (\id\oti\id\oti\coP)(\coa) =
  (\coa\otim\UN)\cdot(\id\oti\coP\oti\id)(\coa) \cdot(\UN\otim\coa)\,;
  \labl{pen}
\itE and the hexagon identities
  \begin{eqnarray} &&\coa_{231}\cdot(\coP\oti\id)(\coc)\cdot\coa_{123}
  =\coc_{13}\cdot\coa_{132}\cdot\coc_{23} \,, \label{hex1} \\[1 mm]
  &&\coa_{312}^*\cdot(\id\oti\coP)(\coc)\cdot\coa_{123}^*=\coc_{13}\cdot
  \coa_{213}^*\cdot\coc_{12}\, .  \label{hex2} \end{eqnarray}
(Here I employ the common notation to write $\coa_{231}\simeq\coa^{(2)}
\otim\coa^{(3)} \otim\coa^{(1)}$, $\coc_{13}\simeq\coc^{(1)}
\otim\UN\otim\coc^{(2)}$, etc.\ for $\coa\equiv\coa_{123}\simeq
\coa^{(1)}\otim\coa^{(2)}\otim\coa^{(3)}$ etc.)

\ite The tensor product is modular in the sense
that the so-called {\em monodromy matrix\/} which is an
element of $M_{\Sigma 
 _p n_p}(\h)$ (see \erf{yrs} below) must be invertible.
\end{itemize}

\noindent To summarize: The \qs\ \h\ of a rational field  theory
is a \findim\ unital associative \stalg\ having a coproduct,
co-unit and antipode; the coproduct is quasi-cocommutative and
quasi-coassociative, the \repcat\ of \h\ is
a braided monoidal \cst-category, and the monodromy matrix is invertible.
Such structures have been given the name {\em \rha s}; they were
first considered by Vecserny\'es \cite{vecs} and Szlach\'anyi
\cite{szla,szve3}, modifying and extending earlier ideas of
Mack and Schomerus \cite{masc3,jf23}.
\vskip 2mm

As the coproduct is not necessarily coassociative and unit preserving,
\rha s are generically not genuine Hopf algebras.
But they share a lot of the properties of quasitriangular
quasi Hopf algebras \cite{drin7} and weak quasi Hopf algebras \cite{masc3}.
The main distinctive features are the ${}^*$-algebra properties and the
further restriction that the monodromy matrix be invertible.
It is worth mentioning that in early treatments \cite{mose3,Algs} of \qss
in \cft\ it was shown that the symmetry algebra is endowed with
a coproduct. In addition, however, it was {\em assumed},
for no particular reason, that this coproduct be coassociative,
and consequently the results of these investigations do not
describe the most general case.

If \cop\ is coassociative, i.e.\ $\coa=e\otim e\otim e$, then the
relation \erf{eid} implies that \cop\ is unital.
On the other hand \cop\ is necessarily non-unital, i.e.
$\cop(e)\neq e\otim e$, if an integer-valued dimension function does
not exist. For an algebraist, the non-unitality is certainly not the most
natural property, but it is the only way in which in the absence of an
integer-valued dimension function the axioms of a \rha\ can be satisfied.
Namely, just evaluate $e$ in
the product \rep\ $\hrep_p\oti\hrep_q$, either directly which yields
$(\hrep_p\otim\hrep_q)(\coP(e))$, or after application of the tensor
product decomposition which yields
  \be  \loplus_r \npqr\hrep_r(e)= \loplus_r \npqr\one_{n_r} \,. \ee
If \cop\ is unital, the first expression evaluates to
  \be  (\hrep_p\oti\hrep_q)(\coP(e))=\hrep_p(e)\otim\hrep_q(e)=\one_{n_p}
  \otim\one_{n_q} \,, \ee
and hence equality of the expressions implies that the set $\{n_p\}$ of
integers satisfies the defining relation \erF{dpqr} of a dimension
function.

\section{Four space-time dimensions}

Because of the importance of {\em local\/} observable algebras,
the different topological nature of causal complements suggests that
there is a major distinction between ${\cal D}\geq4$ and ${\cal D}<4$
space-time dimensions.
This is indeed the case. While in low-dimensional space-time, the
\suse s generically carry \rep s of the braid groups \B n,
for ${\cal D}\geq4$ one is dealing exclusively with \rep s of
the permutation groups \Sy n, so that in particular the \statdim s
are integral. Further, it turns out \cite{doro}
that in space-time dimension $\geq4$ the integral statistical
dimension $d_r$ plays simultaneously also the role of the dimension
$n_r$ of the \rep\ $\hrep_r$ of the \qs\ and of the multiplicity of
the corresponding sector $\calh_r$ in the \hilsph, that is
  \be  d_r=n_r  \,.  \ee

In the framework of \DHR sectors,
it is even possible to reconstruct the field \alg\ \calf\ from
the observable \alg\ and its \susE structure, in such a manner
that the charged fields generate the sectors by action on vectors
in the vacuum sector. Let me mention some details of this
\doro \cite{doro,doro3} construction. One can prove that
for space-time dimension $\geq4$ the following holds:
\begin{enumerate}
\item First, with $f^i_p$ standing for a field in the sector
labelled by $p$, in sloppy notation the relation
  \be  f_p^i\,f_q^j=\sum_{k=1}^{n_q}\sum_{l=1}^{n_p}
  (\rR_{(p,q)})_{ij}^{\ kl}\,f_q^k\,f_p^l\,, \ee
holds with numerical $n_pn_q\times n_pn_q$-matrices $\rR_{(p,q)}$.
\item The sector labels $p$ correspond to the \findim\ \irrep s\ of a
compact group $G$, and the labels $i$ correspond to a basis of the
associated $G$-modules; in particular, the \statdim s $d_p$ are given by
the dimensionalities $n_p$ of these \irrep s.
The \qs\ \h\ is the group algebra of $G$, $\h\cong\complex G$.
$G$ plays the role of a global gauge group;
if it is a Lie group, then in a Lagrangian formulation
its Lie \alg\ can be interpreted as
the \alg\ of conserved Noether charges \wrt the \sym.
\item The tensor product of $G$-\rep s is isomorphic to the product of sectors.
\item Up to an overall plus or minus sign, corresponding to the
distinction between bosons and fermions, the matrices $\rR_{(p,q)}$
are products of \cgc s which correspond to the composition of
intertwining maps\, $\hrep_p\times\hrep_q\to \loplus_r \hrep_r
\leftarrow \hrep_q\times \hrep_p.$
 \end{enumerate}
In short, the category of DHR \rep s is isomorphic to the \repcat\
of some compact group $G$, and apart from the multiplicities corresponding to
these $G$-\rep s, the sectors are either bosonic or fermionic.
In terms of particles, the \doro result means that parabosons and
parafermions are equivalent to ordinary bosons and fermions,
respectively, that carry a nontrivial \rep\ of a compact internal
symmetry group; in simple Lagrangian models of parabosons or -fermions,
this equivalence is realized in terms of a Klein transformation of the
fields. Note that according to the construction, the number of sectors is
necessarily infinite. This does not imply, however, that there is an
infinite number of particles which are elementary in the sense that in
a path integral framework they correspond to elementary fields in the
Lagrangian; rather, in this framework the sectors would typically
correspond to multi-particle states or bound states.

In the proof of the \doro results, use is made of
the \takrein reconstruction theorem which states that to
any symmetric monoidal \cat\ (i.e., monoidal category for which the
commutativity isomorphism squares to the identity)
for which a functor to the \cat\ of
\findim\ vector spaces exists, one can construct a group whose
\repcat\ is equivalent to the original \cat.
To perform this \takrein reconstruction, one must know the intertwiners
between isomorphic \rep s (in particular for the decomposition of
tensor products)
rather explicitly. Therefore it is not directly applicable to \aft,
where only the reference endomorphisms $\rho_p$ but not the
\rep s $\pi_p$ are known well enough. Rather, the appropriate version of
the reconstruction theorem is the one due to Deligne, where the
requirement of the existence of the functor to \findim\ vector spaces
is replaced by the axiom that a
certain number, the so-called rank of an object of the \cat,
is always integral. In the \doro construction, this rank is
given by the \statdim\ $d_p$;
in particular, the map that associates $d_p$
to the sector $p$ is an integer-valued dimension function.

In terms of the \rep\ theory of \vna s, the \doro result is
highly non-trivial: it shows that the multiplication law for physical
\rep s of these complicated \alg s just mimicks that for
compact groups, a fact that certainly would not be expected a priori.
This non-triviality becomes particularly obvious when one tries
to generalize the construction to low space-time dimensions.
In that case, the \statdim s are typically no longer integral,
so that the \doro arguments break down already at an early stage.
It is thus an open question whether there
exists an analogue of the construction for the case of \rha s.

\section{Truncated quantum groups}

Let me return now to the case of rational field theories, and hence
to low space-time dimensionality.
As already mentioned, an alternative approach to \qs\ is via
weak quasi Hopf algebras \cite{masc3}. These, in turn, can often
be obtained by a suitable truncation procedure from an associated
\qg\ ${\cal U}={\sf U}_q(g)$,
where $g$ is a semisimple Lie \alg\ and the deformation parameter
$q$ is a primitive root of unity.

The \rep\ theory of $\cal U$ for $q$ a root of unity differs significantly
from the one for generic $q$ (where it is isomorphic to the \rep\
theory of $g$). As a consequence, one must truncate the \repcat\
\cite{fugp2,Algs} by restricting the set of allowed \rep s to the
physical, i.e.\ unitarizable
ones, and removing all non-unitarizable sub-\rep s from the tensor
products of unitarizable ones (this is consistent because the
non-unitarizable \rep s generate an ideal of the \repcat).
But for a field theoretic description, this not yet enough; rather,
to avoid inconsistencies one must \cite{masc2}
perform the truncation already at the level of the algebra, namely
by defining the \qs\ as the quotient \,$\ut:={\cal U}/{\cal J}$
of \,$\cal U$ by the ideal
  \be  {\cal J}=\{u\in{\cal U}\mid \rphys(u)=0\} \,, \ee
where \rphys\ denotes the direct sum of all physical \rep s.
The quotient \ut\ is a weak quasi Hopf algebra, and hence satisfies
most of the defining properties of a \rha; whether \ut\ or some similar
object can be endowed
with the full structure of a \rha, including in particular all
properties related to the $^*$-involution, is not known.

It must be emphasized that, while the \qg\ $\cal U$ serves as a
starting point of the construction, in the truncation process
actually most of its structure gets lost.
In particular, typically various distinct \qg s lead, by
truncation, to equivalent weak quasi Hopf \alg s \cite{jf23}.
The equivalence of these \alg s is however not provided by ordinary
isomorphisms, as typically the dimensionalities of different such
\alg s do not coincide. This happens because
by construction the numbers $n_p$ are just given by with the
dimensionalities of the modules $L_p(g)$ of the
underlying semisimple Lie \alg\ $g$; for instance, $n_p=p+1$
for $p=0,1,...\,,k$ (with $k$ the level of the associated
untwisted affine Lie \alg) if $g={\rm sl}_2$.

On the other hand, an advantage of this construction is that
it allows for employing the well-developped
(see e.g.\ \cite{FRke,FUch,CHpr}) theory of unitarizable \rep s
of quantum groups at roots of unity.
For example, the coproduct of \ut\ is of the form
  \be  \cop([u])=[{\cal P}\cdot\coP_{\cal U}(u)] \,, \ee
where $\coP_{\cal U}$ is the coproduct of $\cal U$ and $\cal P$
a projector in ${\cal U}\oti{\cal U}$. Moreover,
whenever no fusion coefficient \npqr\ is larger than 1 (and hence
e.g.\ for $g={\rm sl}_2$ \cite{masc3} and for arbitrary $g$ at
level one \cite{jf23}), there is a simple formula for $\cal P$,
namely
  \be  {\cal P}=\sum_{p,q,r}\npqr\,({\cal P}_p\otim{\cal P}_q)\cdot
  \cop({\cal P}_r) \,, \ee
with ${\cal P}_q\in{\cal U}$ a set of independent projectors
spanning the center of \,$\cal U$ (in \rphys).

\section{Reconstruction}

As it turns out, it is possible to reconstruct from the \qs\ \h\
various aspects of the field theory. In particular there is a
construction analogous to the one that associates the statistics
parameter, and hence the numbers $d_p$ and $\oeta_p$, to a
\rep\ $\pi_p$ of the observable \alg\ \ca.
{}From the list of properties of \h\ presented in section 4, it is far
from obvious what this construction could be in the \findim\ setting
of the \qs\ of a rational field theory.
The clue to the construction is the use of so-called
{\em amplimorphisms}, i.e.\ $^*$-monomorphisms $\mU_p$ from \h\ to
matrix \alg s over \h,
  \be  \mU_p:\quad \h\rightarrow M_{n_p}^{}(\h) \,, \labl{amp}
instead of representations of \H, which are ${}^*$-algebra homomorphisms
$\hrep\colon\ \H\to M_n(\complex)$.
Note that to be able to produce non-integral statistical dimensions $d_p$,
one has to employ amplimorphisms of \h\ rather than just endomorphisms.
It is a remarkable result that one can indeed arrive at
non-integral statistical dimensions, and even
non-integral embedding indices $d_p^2$, by \findim\ constructions.

Before treating the amplimorphisms in some detail, let me say a few
words about the \rep s \hrep. They are unitarizable and completely reducible;
the defining irreducible representations $\hrep_p$ read
  $ \h\ni a\equiv\sum_p \sum_{i,j} a_p^{ij}\e pij \mapsto
  (\hrep_r(a))^{ij}=a_r^{ij}$.
Here $\e rij,\; i,j=1,2,\ldots ,n_r$, denote the matrix units
which provide a basis of \h.
In this basis the $^*$-involution reads simply $(\e pij)^* =\e pji$.

With the help of the coproduct, one defines the product of representations
as $(\hrep_1\tims \hrep_2)(a):=(\hrep_1\oti\hrep_2)(\coP(a))$.
The properties \erf{coc1}, \erf{coa1} etc.\ of \coc\ and \coa\
ensure the commutativity and associativity
of the product of representations up to unitary equivalence.
In the basis of matrix units, the co-unit obeys $\cou(\e pij)=1$ for
$\e pij=e_0$, $\cou(\e pij)=0$ else; as a consequence,
products of representations with the co-unit \cou\ lead to representations
that are equivalent to the original ones.

The antipode can be chosen such that $\apo(\e pij)=\e {p\cj}ji$.
Then the conjugate $\barhrep$ of a representation $\hrep$ satisfies
$(\barhrep_p(a))^{ij}:=(\hrep_p(\apo(a)))^{ji}=(\hrep_{p\cj}(a))^{ij}$
for all $a\in\H$.
The properties of \apo\ ensure that $\barhrep\colon\,\H\to M_n(\H)$ is a
${}^*$-homomorphism, with $n$ the dimension of $\hrep$.

Amplimorphisms contain the full information about \rep s. Namely, on one
hand, any amplimorphism $\mu\colon H\to M_m(H)$ induces a representation
by composing it with the co-unit,
  \be  \hrep_\mu:=\cou\circ\mu\colon\ \H\to M_m(\Co),\qquad
  \hrep^{ij}_\mu(a):=\cou(\mu^{ij}(a))  \ee
for $a\in\H$ and $i,j=1,2,...\,,m.$ On the other hand,
any (non-zero) representation $\hrep$ of \H\ defines an amplimorphism
$\muhrep\colon H\to M_m(H)$, with $m$ the dimension of $\hrep$, through
  \be  \muhrep:=(\id\oti\hrep)\circ\cop \,; \ee
amplimorphisms of this type are called a {\em special amplimorphisms}.
Similarly,
an amplimorphism $\nu\colon\H\to M_n(\H)$ is called {\em natural\/}
if $\nu\sim\muhrep$, i.e.\ if there is an equivalence $T\in(\muhrep^{}
\vert\nu)$ for some representation $\hrep\colon\H\to M_n(\Co)$.
(Amplimorphisms $\mu$ and $\nu$ are called equivalent, $\mu\sim\nu$, if
there is a $T\in(\mu|\nu)$ with $TT^*=\mu(\UN),\;T^*T=\nu(\UN).$
The space $(\mu\vert\nu)$ of intertwiners between
$\mu\colon\H\to M_m(\H)$ and $\nu\colon \H\to M_n(\H)$ consists of all
$T\in M_{m\times n}(\H)$ such that $\mu(a)T=T\nu(a)$ for all $a\in\H$,
and $\mu(\UN)T=T=T\nu(\UN)$.)
One can define subobjects, direct sums and an associative product of
amplimorphisms; the latter reads $(\mu\times\nu)^{i_1j_1,i_2j_2}(a)
=\mu^{i_1i_2}(\nu^{j_1j_2}(a))$.

To see the relation with the braid group, combine the identity
  $\R_{12}\cdot(\coP\oti\id)(\R)=(\copP\oti\id)(\R)\cdot\R_{12}$
with the first hexagon equation. This leads to
the {\em quasi Yang\hy Baxter equation\/}
  \be  \R_{12}^{}\Q^*_{231}\R_{13}^{}\Q_{132}^{}\R_{23}^{}\Q^*=
  \Q^*_{321}\R_{23}^{}\Q_{312}^{}\R_{13}^{}\Q^*_{213}\R_{12}^{} \,. \labl{qybe}
Now define the maps
  \be  \sigma_i:=\left\{ \begin{array}{ll} \RR_{i,i+1} & {\rm for}\
  i\in2\zet, \\[1 mm] \QQ_{i+2,i+1,i}\circ\RR_{i,i+1}\circ\QQ^*_{i+2,i+1,i}
  & {\rm for}\ i\in2\zet+1, \end{array} \right. \labl{sigpq}
acting on the infinite tensor product $\h\oti\h\oti\h\oti
\ldots$\,, where $\QQ_{ijk}$ stands for multiplication with $\Q_{ijk}$
from the right, and $\RR_{ij}$ for multiplication with $\R_{ij}$ followed
by transposition $\tau_{i,j}$ of the indicated tensor
factors. These maps $\sigma_i$ satisfy the defining relations
of the infinite braid group \B\infty\ and hence furnish a \rep\ of
\B\infty\ on $\h\oti\h\oti\h \oti\ldots$\,\,.

Further information about these braid group \rep s is obtained with
the help of the amplimorphisms of \H.
The braiding properties of amplimorphisms are encoded in the {\em
statistics operators\/} \veps\
which are intertwiners between $\mu\times\nu$ and $\nu\times\mu$.
For special amplimorphisms $\mu_p,\, \mu_q$ corresponding to
representations $\hrep_p$ and $\hrep_q$, \veps\ is defined by
  \be  \vept (\mu_p;\mu_q)=[(\id\oti \hrep_q\oti \hrep_p)(\coa) ]\cdot
  [(\id\oti \tau)\circ(\id\oti \hrep_p\oti \hrep_q)(\coc)]\cdot
  [(\id\oti \hrep_p\oti \hrep_q)(\coa^*)] \,. \ee
\vept \ is unitary in the sense that
  $\vept (\mu_p;\mu_q)\cdot\vept (\mu_p;\mu_q)^*=
  (\mu_q\times\mu_p)(e)$ and
  $\vept (\mu_p;\mu_q)^*\cdot\vept (\mu_p;\mu_q)=
  (\mu_p\times\mu_q)(e)$,
and upon multiplication the statistics operators give rise to a \rep\
of coloured braids.

A {\em left inverse\/} $\Phi_\nu\colon\, M_n(\H)\to \H$ of an amplimorphism
$\nu\colon\, \H\to M_n(\H)$
is a positive linear map which satisfies $\Phi_\nu(\one_n)=\one$ and
 $\Phi_\nu(\nu(a)\cdot B\cdot\nu(c))=a\cdot\Phi_\nu(B)\cdot c$
 for all $a,c\in\H$ and all $B\in M_n(\H)$. For special amplimorphisms
$\mu_p$, it can be defined as
  \be  \Phi_p(A):=P^*_p\cdot\bar \mu_{p\cj}(A)\cdot P_p \qquad
  \mbox{for all }\ A\in M_{n_p}(\H) \,, \ee
where
  \be  P_p^{ij,\cdot}=[{\rm tr}\,\hrep_p(\br\br^*)]^{-1/2}_{}\,\coa^{(1)}
  \cdot (\hrep_p)^{ji}(\coa^{(3)}r^*\apo(\coa^{(2)})) \ee
for $i,j=1,2,...\,,n_p$, which is a partial isometry in $(\mu_{p\cj}\times
\mu_p\vert\id)$.

Given the statistics operators and left inverses, one can finally
compute the {\em statistical parameter\/} $\lambda_p\in\H$ of a special
amplimorphism $\mu_p$ as
  \be  \lambda_p=\Phi_p\circ\Phi_p(\veps(\mu_p;\mu_p)) \,. \ee
Similarly, the monodromy matrix $Y$ is
  \be  Y_{rs} =d_rd_s\cdot\Phi_r\circ\Phi_s(\veps(\mu_r;\mu_s)\cdot
  \veps(\mu_s;\mu_r)) \,.  \labl{yrs}
It can be shown that
  \be  \lambda_p=(\oeta_p/d_p)\,\cdot \UN  \ee
with $\oeta_p$ the statistics phase and $d_p$ the statistical dimension
of the \suse\ $\pi_p$, and that
$Y_{rs}=y_{rs}\cdot\UN$ with $y_{rs}\in\Co$. Further,
if $Y$ is invertible, then by defining
  \be  V({\rm S})_{rs}:=\vert\sigma\vert^{-1}\cdot y_{rs},\qquad
  V({\rm T})_{rs}:=(\sigma/\vert\sigma\vert)^{1/3}\cdot
  \delta_{rs}\,\oeta_s \ee
with $\sigma\equiv\sum_p d_p^2/\omega_p$, one obtains a unitary \rep\
$V$ of the double cover SL$(2,\zet)$ of the modular group.
In the case of \cfts, this \rep\ fixes the value of the Virasoro
central charge modulo 8.

\section{Classification}

Having identified \rha s as the \qss of rational field theories, there
remain two basic tasks: the classification of such algebras, and the
identification of the \rha s relevant for specific theories.

The coproduct \cop\ of \h\ can be viewed as an
embedding of \H\ into $\H\oti\H$. As such embeddings are relevant
only up to inner unitary automorphisms of $\H\oti\H$, there are
equivalences with appropriate elements $U\in\h\oti\h$.
This `gauge freedom' can be exploited to present a \rha\ in a
canonical form; in particular, with a suitable choice of $U$ one
can put $a_{\rm r}=e=a_{\rm l}$, as was already used above,
and \br\ and \bl\ can be taken as invertible central elements
of $\H$ one of which is positive.
As might have been expected, even after fixing this gauge freedom a
complete classification of \rha s is extremely difficult; actually
at present it is beyond reach. The classification involves in particular
the classification of solutions to the pentagon  and hexagon equations
\erf{pen} -- \erf{hex2}.

However \cite{vecs,fugv3},
for any given (finite) number of sectors and any fixed set of
\furu s among them, it is in principle straightforward to
write down the most general compatible coproduct, solve the pentagon
and hexagon identities, and compute the statistics operators, left
inverses, statistics parameters, and the monodromy matrix.
In particular, after suitable choice of `coordinates' in the
space of multi-matrix algebras, one can encode the necessary
manipulations in a computer program. At least for small numbers of
sectors, the solutions can then also be obtained in practice
\cite{fugv3}. That a complete solution of the pentagon
and hexagon equations, which constitute a huge system of coupled
nonlinear equations in the relevant coordinates, is indeed possible for
non-trivial \furu s, is related to the fact that this system of
equations is actually highly over-constrained.

What can so far {\em not\/} be obtained in this approach to
the classification are the integers $n_p$, and hence the
dimensionalities of the \rha s. Rather, in the algorithm just
mentioned an important ingredient is that the coordinates are
chosen in such a manner that they do not depend on these integers
at all; to any solution to the remaining constraints there is then
associated an infinity of possible choices of the $n_p$,
which is only restricted by the conditions $n_0=1$ and $n_{p\cj}=n_p$
and by the inequality \erf{npqr}. For any set of prescribed \furu s, there
exists a minimal choice of these integers, but at present no physical
argument is known which would exclude any non-minimal choice.
For the minimal choice, one has in particular $n_p=1$ whenever
$d_p=1$, and
moreover $d_p=d_q$ if $\pi_p\cong\pi_q\times\pi_r$ for some $r$
with $d_r=1$ (in \cft\ terms, this means that $n_p=1$ for simple currents,
and $d_p=d_q$ if
the sectors $p$ and $q$ lie on the same simple current orbit).
In the case of \tqg s, the dimensionalities $n_p$ are always
non-minimal; in particular $n_p$ is larger than 1
for all simple sectors different from the vacuum sector.

It is also straightforward to associate to a \rha\ obtained via this
classification a (unitary) \cft; namely, one just has to search the
set of known \cfts\ for those which have the correct number of primary
fields and the correct fusion rules, conformal dimensions and
Virasoro central charge. For example, for the Ising \furu s \cite{vecs}, one
obtains the Ising model as well as the level 2 $A_1$ and $E_8$ and the
level 1 $B_r$ Wess\hy Zu\-mi\-no\hy Witten theories. Thereby
the classification of rational Hopf algebras corresponds to a partial
classification of unitary rational \cfts, namely, the possible fusion rules,
the conformal dimensions modulo integers, and the Virasoro central charge
modulo 8. It is not yet clear whether different \cfts\ in this
classification that share the same \rha\ (e.g.\ in the example of
the Ising \furu s, the level 1 $B_r$ theories with $r$ differing by
an integer multiple of 8) are to be considered as distinct \qfts.

To conclude, let me mention that for all \rha s analyzed so far
\cite{fugv3} it turned out to be possible to
identify associated unitary rational \cfts. If this is a generic
feature, it implies that as far as it is only the \rha\ that matters,
{\em any\/} \twodim\ rational \qft\ can be viewed as being
equivalent to a rational {\em conformal\/} field theory.

   \newcommand{\wb}{\,\linebreak[0]} \def\wB {$\,$\wb}
   \newcommand{\J}[1]     {{{#1}}\vyp}
   \newcommand{\Jj}[1]    {{{#1}}\vyP}
   \newcommand{\JJ}[1]    {{{#1}}\vyp}
   \newcommand{\bi}[1]    {\bibitem{#1}}
   \newcommand{\Bi}[1]    {\bibitem{#1}}
   \newcommand{\PRep}[2]  {preprint {#1}}
   \newcommand{\PhD}[2]   {Ph.D.\ thesis (#1)}
   \newcommand{\Erra}[3]  {\,[{\em ibid.}\ {#1} ({#2}) {#3}, {\em Erratum}]}
   \newcommand{\BOOK}[4]  {{\em #1\/} ({#2}, {#3} {#4})}
   \newcommand{\inBO}[7]  {in:\ {\em #1}, {#2}\ ({#3}, {#4} {#5}), p.\ {#6}}
   \newcommand{\inBOnoeds}[6]  {in:\ {\em #1} ({#2}, {#3} {#4}), p.\ {#5}}
   \newcommand{\vyp}[4]   {\ {#1} ({#2}) {#3}}
   \newcommand{\vyP}[3]   {\ ({#1}) {#2}}
   \newcommand{\vypf}[5]  {\ {#1} [FS{#2}] ({#3}) {#4}}

   \def\acam  {Acta\wB Appl.\wb Math.}
   \def\acma  {Acta\wB Math.}
   \def\adma  {Adv.\wb Math.}
   \def\anma  {Ann.\wb Math.}
   \def\anop  {Ann.\wb Phys.}
   \def\bams  {Bull.\wb Amer.\wb Math.\wb Soc.}
   \def\blms  {Bull.\wB London\wB Math.\wb Soc.}
   \def\bsbm  {Bol.\wb Soc.\wb Bras.\wb Math.}
   \def\bsmf  {Bull.\wb Soc.\wb Math.\wB de\wB France}
   \def\busm  {Bull.\wb Sci.\wb Math.}
   \def\coia  {Com\-mun.\wB in\wB Algebra}
   \def\coma  {Con\-temp.\wb Math.}
   \def\comp  {Com\-mun.\wb Math.\wb Phys.}
   \def\cpma  {Com\-pos.\wb Math.}
   \def\crap  {C.\wb R.\wb Acad.\wb Sci.\wB Paris}
   \def\foph  {Fortschr.\wb Phys.}
   \def\fuaa  {Funct.\wb Anal.\wb Appl.}
   \def\ijmb  {Int.\wb J.\wb Mod.\wb Phys.\ B}
   \def\ijmc  {Int.\wb J.\wb Mod.\wb Phys.\ C}
   \def\ijmp  {Int.\wb J.\wb Mod.\wb Phys.\ A}
   \def\inma  {Invent.\wb math.}
   \def\jams  {J.\wb Amer.\wb Math.\wb Soc.}
   \def\jgap  {J.\wb Geom.\wB and\wB Phys.}
   \def\joal  {J.\wB Al\-ge\-bra}
   \def\jodg  {J.\wb Diff.\wb Geom.}
   \def\jofa  {J.\wb Funct.\wb Anal.}
   \def\jomp  {J.\wb Math.\wb Phys.}
   \newcommand{\kniz}[2] {\inBO{The Physics and Mathematics of Strings,
              Memorial Volume for V.G.\ Knizhnik} {L.\ Brink et al., eds.}
              \WS\Si{1990} {{#1}}{{#2}} }
   \def\lemp  {Lett.\wb Math.\wb Phys.}
   \def\lenc  {Lett.\wB Nuovo\wB Cim.}
   \def\leni  {Lenin\-grad\wB Math.\wb J.}
   \def\liam  {Lectures\wB in\wB Applied\wB Math.}
   \def\maan  {Math.\wb Annal.}
   \def\mams  {Memoirs\wB Amer.\wb Math.\wb Soc.}   
   \newcommand{\mapx}[2] {\inBO{Mathematical Physics X}
              {K.\ Schm\"udgen, ed.} \SV\Be{1992} {{#1}}{{#2}} }
   \def\npbf  {Nucl.\wb Phys.\ B\vypf}
   \def\npbp  {Nucl.\wb Phys.\ B (Proc.\wb Suppl.)}
   \def\nupb  {Nucl.\wb Phys.\ B}
   \def\paaa  {Proc.\wb Amer.\wb Acad.\wB Arts\wB Sci.}
   \def\pams  {Proc.\wb Amer.\wb Math.\wb Soc.}
   \def\pcps  {Proc.\wB Cam\-bridge\wB Philos.\wb Soc.}
   \def\phlb  {Phys.\wb Lett.\ B}
   \def\pkna  {Proc.\wb Kon.\wb Ned.\wb Akad.\wb Wetensch.}
   \def\plms  {Proc.\wB Lon\-don\wB Math.\wb Soc.}
   \def\pnas  {Proc.\wb Natl.\wb Acad.\wb Sci.\wb USA}
   \def\prsa  {Proc.\wb Roy.\wb Soc.\wB Ser.$\,$A}
   \def\prtp  {Progr.\wb Theor.\wb Phys.}
   \def\pspm  {Proc.\wb Symp.\wB Pure\wB Math.}
   \def\ptps  {Progr.\wb Theor.\wb Phys.\wb Suppl.}
   \def\ptrs  {Phil.\wb Trans.\wb Roy.\wb Soc.\wB Lon\-don}
   \def\rpmp  {Rep.\wb Math.\wb Phys.}
   \def\rmap  {Rev.\wb Math.\wb Phys.}
   \def\slnm  {Sprin\-ger Lecture Notes in Mathematics}
   \def\slnp  {Sprin\-ger Lecture Notes in Physics}
   \newcommand{\Suse} [2] {\inBO{The Algebraic Theory of Superselection
              Sectors.\ Introduction and Recent Results} {D. Kastler,
              ed.} \WS\Si{1990} {{#1}}{{#2}} }
   \def\tams  {Trans.\wb Amer.\wb Math.\wb Soc.}
   \def\thmp  {Theor.\wb Math.\wb Phys.}
   \def\AMS    {{American Mathematical Society}}
   \def\AP     {{Academic Press}}
   \def\AW     {{Addi\-son\hy Wes\-ley}}
   \def\BC     {{Ben\-jamin\,/\,Cum\-mings}}
   \def\BIR    {{Birk\-h\"au\-ser}}
   \def\CUP    {{Cambridge University Press}}
   \def\CUPC   {{Cambridge University Press}}
   \def\DP     {{Dover Publications}}
   \def\GB     {{Gordon and Breach}}
   \def\JW     {{John Wiley}}
   \def\KLU    {{Kluwer Academic Publishers}}
   \def\MD     {{Marcel Dekker}}
   \def\MGH    {{McGraw\,\hy\,Hill}}
   \def\NH     {{North Holland Publishing Company}}
   \def\OUP    {{Oxford University Press}}
   \def\PL     {{Plenum}}
   \def\PUP    {{Princeton University Press}}
   \def\SV     {{Sprin\-ger Verlag}}
   \def\WI     {{Wiley Interscience}}
   \def\WS     {{World Scientific}}
   \def\Be     {{Berlin}}
   \def\Ca     {{Cambridge}}
   \def\NY     {{New York}}
   \def\pR     {{Princeton}}
   \def\Si     {{Singapore}}

\def\A       {Algebra}
\def\aff     {affine Lie algebra}
\def\alg     {algebra}
\def\Class   {Classification\ }
\def\class   {classification}
\def\Con     {Conformal\ }
\def\con     {conformal\ }
\def\cua     {current algebra}
\def\dimn    {dimension}
\def\emt     {energy-momentum tensor}
\def\enva    {enveloping algebra}
\def\eq      {equa\-tion}
\def\fts     {field theories}
\def\hopf    {Hopf algebra}
\def\ide     {identification}
\def\jf      {J.\ Fuchs}
\def\km      {Kac\hy Moody}
\def\kma     {Kac\hy Moody algebra}
\def\kze     {Knizh\-nik\hy Za\-mo\-lod\-chi\-kov equation}
\def\lie     {Lie algebra}
\def\lmslns  {London Math.\ Soc.\ Lecture Note Series \# }
\def\q       {quantum\ }

\small
\vskip 6mm {\sc Acknowledgement.} The results reported in this paper
are based on collaboration with A.Ch.\ Ganchev and P.\ Vecserny\'es.
I am also grateful to K.\ Blaub\"ar and A.\ Zoomorv for instructive
yarns and comments, and to A.Ch.\ Ganchev, C.\ Schweigert, and
P.\ Vecserny\'es for carefully reading the manuscript.

\end{document}